# Proof of principle of a high-spatial-resolution, resonant-response γ-ray detector for Gamma Resonance Absorption in $^{14}$N

M. Brandis[a,b], M.B. Goldberg[a], D. Vartsky[a], E. Friedman[b], I. Kreslo[c], I. Mardor[a], V. Dangendorf[d], S. Levi[a], I. Mor[a] and D. Bar[a]

[a] *$^1$Nuclear Physics Division, Soreq NRC,*
  *Yavne 81800, Israel*

[b] *$^2$Racah Institute of Physics, The Hebrew University of Jerusalem,*
  *Jerusalem 91904, Israel*

[c] *$^3$Laboratory for High Energy Physics (LHEP), University of Bern,*
  *Bern 3012 Switzerland*

[d] *$^4$Physikalisch-Technisch- Bundesanstalt (PTB),*
  *Braunschweig 38116, Germany*
  *E-mail*: michal_brandis@yahoo.com

ABSTRACT: The development of a mm-spatial-resolution, resonant-response detector based on a micrometric glass capillary array filled with liquid scintillator is described. This detector was developed for Gamma Resonance Absorption (GRA) in $^{14}$N. GRA is an automatic-decision radiographic screening technique that combines high radiation penetration (the probe is a 9.17 MeV gamma ray) with very good sensitivity and specificity to nitrogenous explosives.

Detailed simulation of the detector response to electrons and protons generated by the 9.17 MeV γ-rays was followed by a proof-of principle experiment, using a mixed γ-ray and neutron source. Towards this, a prototype capillary detector was assembled, including the associated filling and readout systems. Simulations and experimental results indeed show that proton tracks are distinguishable from electron tracks at relevant energies, on the basis of a criterion that combines track length and light intensity per unit length.

KEYWORDS: Detection of explosives; Inspection with gamma rays.

**Contents**

## 1. Introduction

GRA is an element-specific, automatic-decision, radiographic-transmission imaging technique, capable of revealing the spatial density distribution of the element in question within inspected objects first proposed by Soreq NRC (SNRC) as a means of detecting explosives in airline baggage via nitrogen-specific radiographic imaging [1,2]. High nitrogen density is characteristic of most explosives but not of commonly transported benign materials; the GRA method relies on this property to distinguish between them. The GRA method is uniquely well-suited to inspecting large, massive objects such as aviation/marine containers, heavy vehicles and rolling stock, the resonant probe being a high energy γ-ray of 9.17 MeV.

The method was successfully taken under FAA sponsorship through several experimental feasibility rounds, including a proof-of-principle laboratory test launched in collaboration with Los Alamos National Laboratory (LANL) (1989) [3], a blind test on aviation baggage aggregates (1993) [4] and a preliminary run on an aviation container (1998) [5].

For detection of nitrogenous explosives via GRA, the inspected object is scanned by a beam of γ-rays, a fraction of which represents on-resonance flux at 9.172 MeV that is capable of being absorbed resonantly by $^{14}$N nuclei. Thus, in addition to the familiar non-resonant atomic processes that attenuate the transmitted flux, such as Compton scattering and pair production, these γ-rays will also undergo a nuclear resonance component of attenuation that is proportional to the density of nitrogen in the line-of-sight between radiation source and detector. By measuring the transmitted γ-rays flux at energies on and off resonance and normalizing appropriately, both the non-resonant and net resonant (total-nitrogen) components of gamma attenuation can be extracted.

The optimal γ-ray source for the nitrogen GRA method is the de-excitation spectrum of the excited $^{14}N^*$ 9.17 MeV level following 1.746 MeV proton capture via the reaction $^{13}C(p,\gamma)^{14}N$. Since the lifetime of the 9.17 MeV level ($5.1 \times 10^{-18}$ s) is very short compared to ion stopping times (typically $\sim 1 \times 10^{-12}$ s) the emission of the γ-ray occurs in-flight following the recoil of the excited $^{14}N$ nucleus, resulting in Doppler-shifting of the γ-ray. At the resonant angle $\theta_R = 80.66°$ with respect to the proton beam, the nuclear recoil energy losses that occur during emission and absorption of the γ-ray by the $^{14}N$ nucleus are precisely compensated by the Doppler-shifted energy component. Since the 9.17 MeV γ-ray yield into $4\pi$ is $\sim 6 \times 10^{-9}$ γ-rays per proton, it is estimated that an operational explosives detection system with realistic throughput based on this reaction will require a 3-5 mA proton accelerator.

For the kinematic reasons mentioned above the resonant flux is confined to a narrow cone around polar angle $\theta_R = \mathbf{80.66°}$ relative to the proton beam forming a slightly curved fan beam. A linear detector array is positioned on the perimeter of this cone as shown in Fig. 1. The inspected object passes through the fan beam and a radiographic image is formed slice by slice..

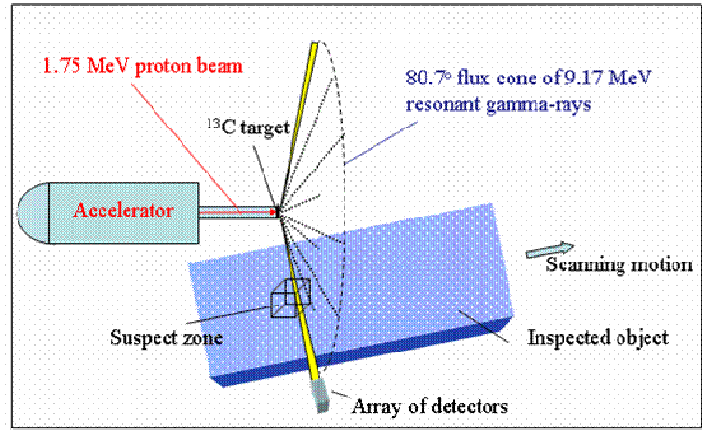

**Fig. 1: Schematic of GRA imaging system showing the 9.17 MeV resonant gamma-ray beam cone. The detector array is positioned on the perimeter of the resonant cone. The inspected object is being moved in a scanning motion between the source and the detector.**

The width Δθ of the resonant cone has been determined by several investigators [8-10]. All experiments consistently show that, the angular aperture size required to include most of the resonant photons is Δθ ~**0.75°**. This aperture corresponds to an energy spread of **520 eV**, which is a factor of ~4 larger than the total intrinsic width of the level $\Gamma_T$=122 eV [10]. The consequence of the emission-line broadening is that only a small intensity fraction within the cone (of order **~0.23**) is on-resonance radiation. Thus, measuring the total intensity of the emission line with a conventional-response detector such as NaI, BGO or Ge, will result in reduced nitrogen contrast sensitivity by this fraction [11].

Soreq NRC has hitherto circumvented this problem by developing nitrogen-rich resonant-response detectors, that are selectively sensitive to photons in the 9.17 MeV region [2,11]. With such detectors the resonant flux is sampled, on an event-by-event basis, by counting 1.5 MeV photo-protons internally-produced via the resonant absorption reaction $^{14}N(\gamma,p)$, the inverse reaction to $^{13}C$ p-capture:

$$\gamma\ (h\nu\ 9.17\ \text{MeV}) + {}^{14}N_{g.s.} \rightarrow {}^{13}C_{g.s}\ (0.12\ \text{MeV}) + p\ (1.5\ \text{MeV})$$

The separation of photo-protons from non-resonant events (electrons and positrons) is realized in currently existing detectors [5] through Pulse Shape Discrimination (PSD) [11,12], a method that utilizes the difference in scintillation light decay time to distinguish electron and proton events. The latter give rise to a somewhat longer signal, spread over a wider time range.

The large angular width of the flux cone required to include most of the resonant photons limits the achievable GRA spatial resolution. Thus the current Soreq resonant detectors shown in Fig. 2 are composed of an Aluminum vessel of $20\times20\times240$ mm$^3$ in dimensions with a rectangular glass window, filled with a special-purpose, nitrogen-rich liquid scintillator.

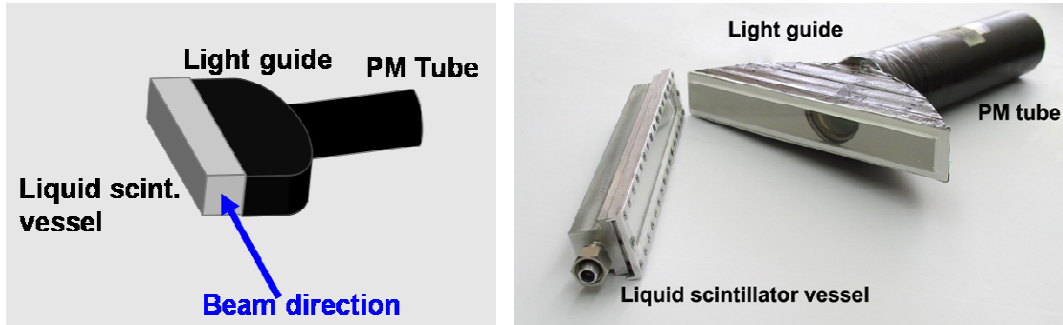

**Fig. 2:. Left: schematic representation of the detector setup. Right: Components of the nitrogen-rich, liquid scintillator resonant detector**

The vessel is coupled via a parabolic Perspex light-guide to a 2" diameter photomultiplier tube. The 20x20 mm$^2$ detector face is presented normal to the γ-ray beam, forming a pixel of these dimensions. The spatial resolution of the GRA system with the present detectors is 5-15 mm, depending on the location of the voxel of interest within the source-detector space. Although Soreq has demonstrated that it is possible to detect thin sheet explosives in cargo using such detectors [5], it would be clearly beneficial to improve the spatial resolution.

In November 2005, within the framework of a DHS/TSA Risk Reduction program Soreq NRC launched an R&D effort aimed at developing novel resonant response detectors with sub-mm resolution, based on imaging the photo-proton and electron-induced tracks in a glass capillary array filled with liquid scintillator [17].

In parallel, an independent approach to high-spatial-resolution resonant detector is developed by the LHEP group, University of Bern, using a Time Projection Chamber filled with a mixture of liquefied Argon and Nitrogen [16]

This paper summarizes the work performed at Soreq NRC so far on the capillary tracking detector and is organized as follows: Sect. 2 describes the concept and the physical processes taking place within the detector. In Sect. 3 a simulation of detector response to 9.17 MeV γ-rays and 1.5 MeV protons is performed. Sect. 4 describes the apparatus, experimental procedures and the preliminary experimental results obtained with the first prototype of such a detector. Sect. 5 provides a short summary.

## 2. Detector principle of operation

The concept of capillary detectors and most of the hardware used in this work are based on the R&D effort performed in CERN by CHORUS and RD46 collaborations [18-20]. Fig. 3 describes the detector concept schematically.

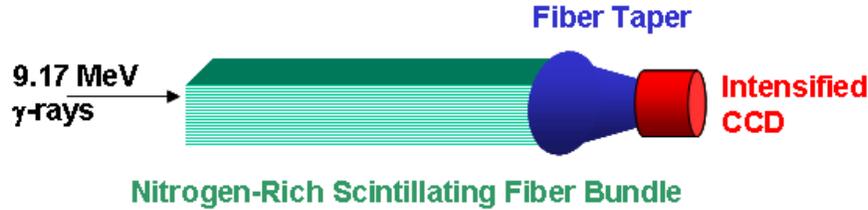

**Fig. 3 Schematic representation of the detector**

The GRA methodology requires that any resonant-response detector, and in particular such a capillary array, should permit separation of resonant interactions from non-resonant interactions on an event-by-event basis [1-4,11].

To this end, the detector consists of a coherent bundle capillary array filled with nitrogen rich liquid scintillator. γ-rays incident on the detector can create photo-protons if they are at the resonant energy. At off-resonance energies, they can create only electrons or electron/positron pairs. The light created in the detector by the particles is transported into an intensified CCD camera which images the projection of particle tracks [17]. Since equal amounts of scintillation light travel along the capillaries in each direction, a good quality reflecting mirror is coupled to one end of the capillary bundle, in order to reflect the light back towards the CCD, thus nearly doubling the amount of light.

Discrimination between particle types produced in the capillary array detector is based on the difference in track projection length and light intensity per unit track length. Light charged particles, such as electrons and positrons, create tracks that are centimeters long while 1.5 MeV protons give rise to much shorter tracks, only ~50 μm long. However, protons leave much more light per unit length along their tracks than electrons. Thus, proton events will give rise to strong bursts of light over a short trajectory of typically a single capillary, which are very distinct from the long, fainter tracks of electrons. Analysis of track projection length and the intensity of light per capillary should permit reliable proton event discrimination with sub-mm spatial resolution.

### 2.1 Track formation by 9.17 MeV γ-rays

The interactions via which the energy is transferred to the scintillator depend on the incident γ-ray energy. Since the special-purpose scintillator inside the capillaries is itself nitrogen rich, resonant γ-rays can undergo the nuclear resonant reaction, producing 1.5 MeV photoprotons and $^{13}$C atoms. The proton path in the scintillator is essentially straight (apart from the extremely rare event of a nuclear collision, when large angle scattering occurs) as the momentum transfer in these ion-ion collisions is relatively small.

Non-resonant γ-ray interactions at ~10 MeV, in organic scintillators are predominantly Compton scattering and, to a lesser extent, pair production. In both interactions the secondary charged particles created are free leptons at various energies. An electron passing through matter dissipates its energy in excitation and ionization, and is scattered in collisions with electrons and nuclei. In its interactions with the matter, the electron may undergo appreciable deflections, so that its path through the scintillator (at energies up to ~10 MeV) is not straight but tortuous.

The primary proton or electron can transfer energy directly to molecule excitation or in a more complicated process, by ionizing a molecule and creating a secondary electron of lower energy that will transfer its energy to molecular excitation. Several generations of such electrons (δ-electrons) can be produced and contribute to the fluorescent excitation process, by which light is emitted in the scintillator.

These secondary and δ-electrons cause track branching and broadening. The branching is more common in primary electron tracks than in proton tracks, whereas the widening is common to both. However, in terms of the measurable quantities, the broadening has an effect on track shape only if, on average, it is larger than a capillary radius.

**2.2 Scintillator light yield and light propagation in capillaries**

The light emission in the scintillator is followed by propagation of this light through the capillaries. The most important figure-of-merit in the capillary detector scintillator is the light output at the end of each capillary. It is dependent on the following parameters: scintillation light yield, the refractive indices of the scintillator and the capillary material, as well as the optical transparency of the scintillator liquid.

The refractive index of the liquid scintillator should be as high as possible relative to that of the capillary material in order for the critical angle of total internal reflection to be as large as possible, allowing a greater portion of the isotropically emitted scintillation light to reach the end of the capillary. This portion, called the trapping fraction, is only a small percentage of the total light yield. For an ideal circular fibre it is equal to $1/2[1-(n_C/n_L)^2]$ [21], where $n_C$ and $n_L$ are the refractive indices of the capillary wall and liquid, respectively. In our experiment we used glass capillaries and a liquid scintillator with refractive indices of 1.49 and 1.62 respectively. In such a configuration the trapping fraction is expected to be up to 7.7% in each direction along the capillary. In practice, capillary imperfections and partial filling may reduce the trapping fraction to about 4-5% [21]. If a good quality reflecting mirror is attached to one end of the capillary array we expect about 10% of the overall created light to be available for the electro-optical readout.

**2.3 Spatial resolution of the track**

Instrumentally, the resolution with which the track can be observed is determined by the diameter of the constituent capillaries. The narrowest capillaries available are 10 μm in diameter; that is therefore the upper limit on the resolution. The glass partition between capillaries needs to be taken into account when considering the spatial resolution. in the experiment conducted at the Laboratory for High-Energy Physics (LHEP), University of Bern, it had a thickness of ~11 μm for the 20 μm capillary diameter array.

In addition there are several physical factors that can cause degradation in spatial resolution: 1) the track properties of the secondary ionizing particles. For electrons, under certain conditions, insufficient energy may be deposited in some of the capillaries to be detected, thus creating an intermittent track. 2) the amount of light traversing through the capillary walls into adjacent capillaries will cause pixel cross-talk. In order to minimize this effect, Extra-Mural-Absorbers (EMU) should be placed between the capillaries and preferably integrated into the capillary matrix. 3) in practice, the spatial resolution is also determined by the optical readout system.

## 3. Simulation of the detector response

The simulations were performed using Geant4 simulation Toolkit, version 4.9.2, distributed by CERN [22,23].

At 9.17 MeV γ-ray energy, the dominant non-resonant interaction process is the Compton Effect, pair production being weak (as the scintillator is low-Z) and photoelectric contributions essentially negligible. All these interactions are included in the simulation application written here. Secondary and tertiary electron interactions in this simulation include: ionization (the principal process inside the scintillator), multiple scattering and Bremsstrahlung radiation. Processes undergone by positrons include annihilation as well.

### 3.1 Simulation detector geometry

Sssc  In order to simulate a capillary array, equi-distant cylinder shaped "incisions" of scintillator-like material were incorporated in a square block of glass-like material, 25×25×350 mm$^3$ in dimensions. Two capillary arrays of different basic cell dimensions were used in the simulation runs (see Table I).

**Table I: Details of different simulated array geometries**

| Variant # | Basic Cell Diameter (µm) | Wall width (µm) | Number of capillaries |
|---|---|---|---|
| 1 | 20 | 11 | 806 x 806 |
| 2 | 50 | 5 | 454 x 454 |

The first variant simulates the capillary array that was available from the CHORUS project (see Sect. 5). Smaller capillaries offer superior tracking resolution, but the amount of light created by the particle per capillary scales with capillary diameter and may thus be insufficient for reliable track identification. This is due to the fact that, below a certain diameter, electrons may deposit insufficient energy in some of the capillaries, thus giving rise to a discontinuous track. The second variant employs capillaries with larger diameters and thinner walls. It represents a much better tradeoff between full tracking capabilities and high spatial resolution.

The simulation was performed in two parts:  1) a pencil beam of 9.17 MeV γ-rays impinged on the central capillary and the non-resonant reaction products (electrons and positrons) were tracked; 2) 1.5 MeV proton source which was disc-shaped and located halfway along the central capillary was simulated. The simulated protons velocity direction was distributed isotropically.  This simulates quite well the photo-protons created by the 9.17 MeV γ-rays that interact resonantly with the nitrogen in the scintillator via the photo-nuclear reaction. Based on the cross-sections for creation of electrons and protons by a monoenergetic 9.17 MeV γ-ray beam [11] the total number of produced electrons is calculated to be 11 times larger than that of protons, for a 35 cm long detector containing 15% nitrogen by weight. This factor has been taken into account in our simulations when evaluating the particle discrimination properties.

For each event the length of track projection (in number of capillaries) is determined and the total energy deposited in the track is calculated. The energy is then converted into number of light photons, taking into account the difference in light yield between electrons and protons, the non linear behaviour of the light yield with proton energy [23] and the fact that the light output of our 15%-nitrogen rich scintillator is reduced by a factor of ~1.6 relatively to a nitrogen free scintillator.

## 3.2 Electron and proton tracks

Example projections of electron tracks in a 20 μm capillary array (Variant I) are presented in Fig. 4. Simulated 9.17 MeV γ-rays impinge on the center of a 25×25 mm² detector. The tracks are color coded as following: red - electrons, blue - positrons, green - γ-rays or X-rays. Fig. 4a shows a case of a Compton scattering followed by production of Bremsstrahlung radiation. Fig. 4b shows a case of pair production followed by an annihilation of the positron. Fig. 4c shows a more complicated case of multiple Compton scattering, Bremsstrahlung radiation and the photoelectric effect.

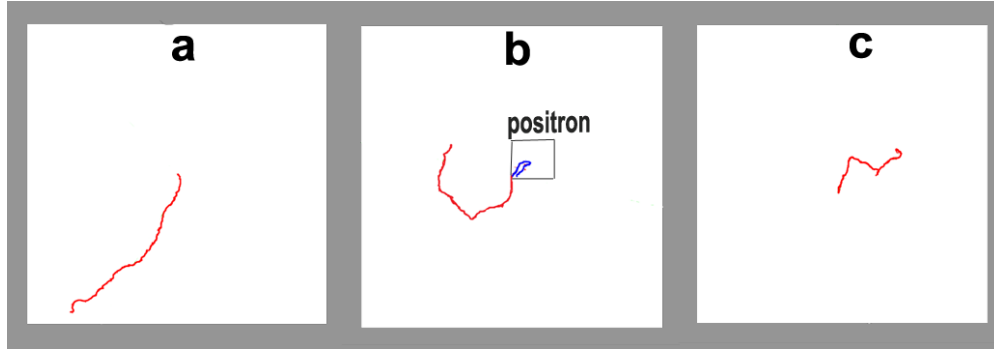

**Fig. 4: Three simulated track projections of electrons and positrons created by 9.17 MeV γ-rays in a 20 μm capillary (Variant I) array, 25×25 mm² in dimensions. The red tracks represent electrons, the blue tracks represent positrons**

Fig. 5 shows a magnified simulated track projection created by a 9.17 MeV □-ray secondary electron (Fig. 5(a)) and a 1.5 MeV proton (Fig. 5(b)) in the capillary array of Variant #1. The length of the proton projection is two capillaries. One can observe the effect of the Bragg distribution, in that the proton deposits most of its energy at the end of the track.

The length of track projection is determined by summing the capillaries along the particle track. Measured track lengths could thus exceed real ones by as much as 1 capillary.

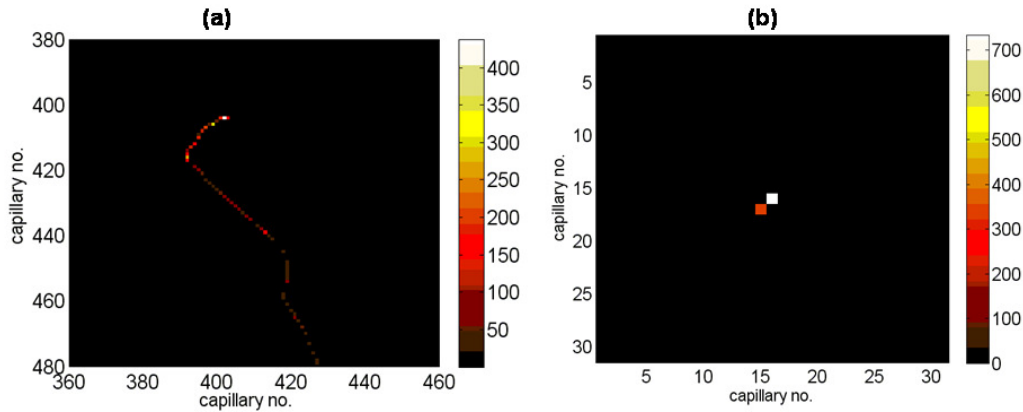

**Fig. 5: Magnified view of simulated tracks in 20 μm (Variant I) capillary array. The color code on the right shows the number of photons/capillary. (a) Part of electron track. (b) 1.5 MeV proton track. Fig 5a and 5b do not have the same magnification.**

Fig. 6 shows the distribution of track length projections (in number of capillaries) of electrons/positrons for the 20 μm diameter capillary. Electron tracks show a wide distribution of lengths with a peak at around 325 capillaries

The protons deposit their energy in up to two capillaries. About 70% of them do so in a single capillary and the remainder - in 2 capillaries.

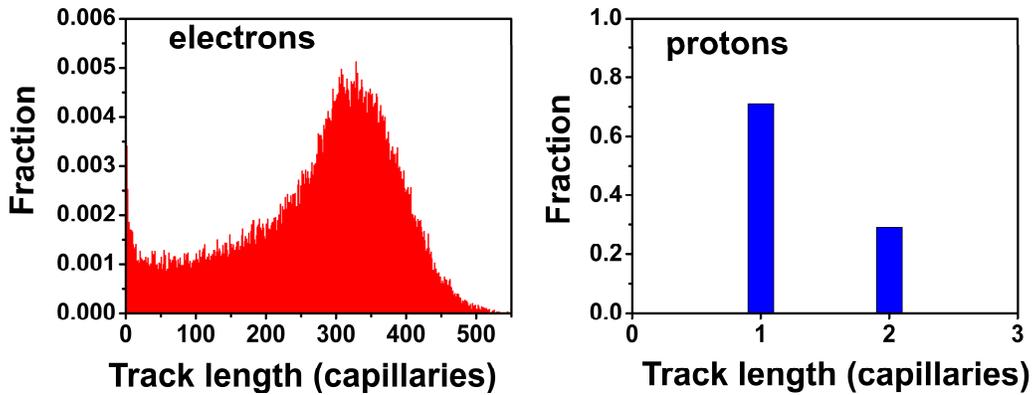

**Fig. 6: Distribution of track length projections (in number of capillaries) for electrons (red) and protons (blue) in Variant I capillary array.**

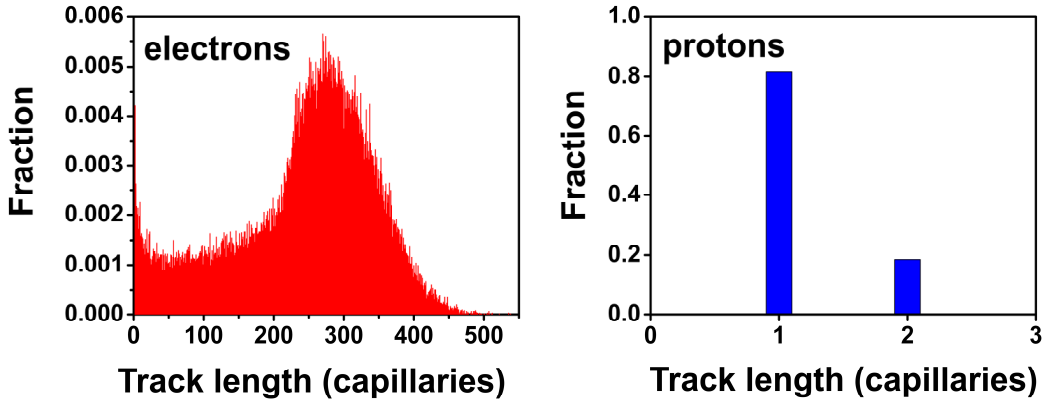

**Fig. 7: Distribution of track length projections (in number of capillaries) for electrons (red) and protons (blue) in Variant II capillary array.**

The Variant II capillary array (50 μm in diameter) shows a similar electron and proton track length (in number of capillaries) distributions (as can be seen in Fig. 7).

Due to the fact that the capillaries are larger, the fraction of a single capillary proton track is slightly higher here.

### 3.3 Track-based particle discrimination

The discrimination between electrons and protons is based on the projection of track length (number of capillaries in track) and the average amount of light per capillary generated by each species. The average amount of light per capillary is derived by calculating the total amount of light in the track and dividing by the number of capillaries in the track projection. In order to calculate the amount of light/capillary that reaches the photocathode of the electro-optical readout we must take into account the light yield of the nitrogen-rich (15% N by weight) scintillator which is ~6100 photons/MeV electron, the difference in light yield between electrons and protons, the non linear behaviour of the light yield with proton energy and the trapping fraction of our capillaries (~10%). Fig. 8 shows the distribution of the average light/capillary on the optical readout's input in Variant I capillaries for electrons (red) and protons (blue). As expected the electrons light/capillary is much lower than for protons.

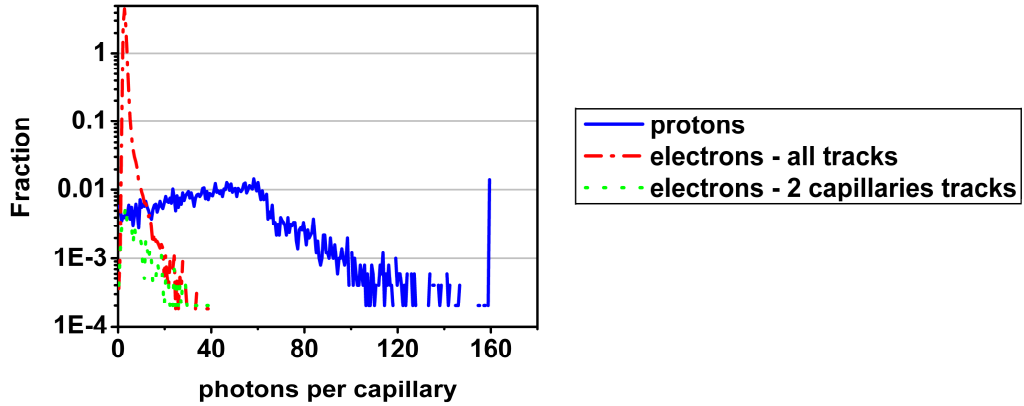

**Fig. 8: Distribution of average light/capillary for 20 μm diameter capillary array. Red-electrons. Blue-protons, Green- electrons with track length of up to 2 capillaries.**

The electron distribution peaks at 3 photons/capillary and drops to a negligible value at around 13 photons/capillary. Only 0.4% of the total number of created electrons are above this value, while 90% of protons are accounted for. One can observe a peak at 160 photons/capillary in the proton distribution. These are events where the proton deposited all its energy in a single capillary. However, there is an additional piece of information that can be invoked for discrimination, namely, the fact that the proton can produce a track length of only up to two capillaries. The green curve shows the fraction of electrons with a track length of up to 2-capillaries, which could be interpreted as protons. These events can be considered as background. When applying this condition together with the 13 photons/capillary threshold, the proton to electron counts ratio is 53. This is a very good signal to background ratio, about a factor of 2 better than that obtained with our present resonant detectors, based on PSD.

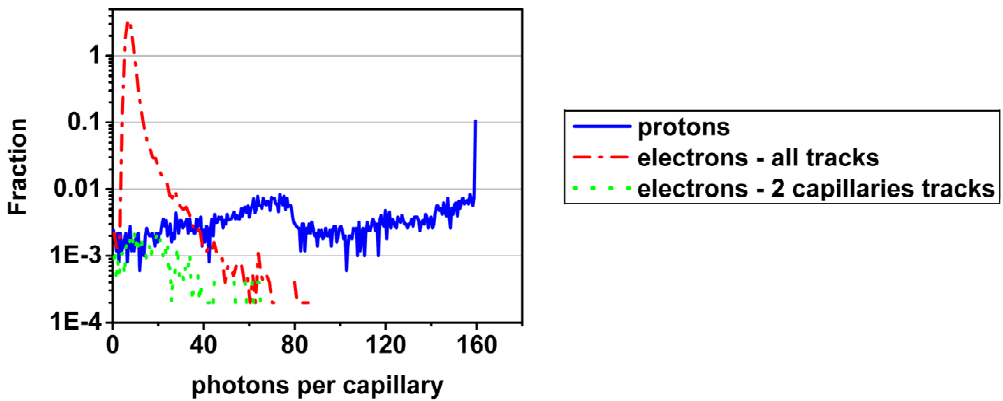

**Fig. 9: Distribution of average light/capillary for 50 μm diameter capillary array. Red - electrons. Blue - protons, Green - electrons with track length of up to 2 capillaries.**

Fig. 9 shows the distribution of the average light/capillary in Variant II capillary array. In this variant, the distribution is shifted towards larger values of light/capillary for both electrons and protons. Here, the electron distribution peaks at about 7 photons/capillary. With a threshold of about 31 photons/capillary the fraction of detected events is 90% and 0.4% for protons and electrons respectively. When the 2-capillaries condition is applied in addition to this threshold the ratio of proton to electron number of events is 56.

In a practical light detector with a quantum efficiency of about 15%, some electrons will form weak and intermittent tracks and will not be visible. Using a binomial distribution to calculate the probability of generating one or more photoelectrons at the photocathode for the light emitted from a capillary we estimate that 38% and 95% of the electron and proton tracks respectively for Variant I capillary array will be observed. Due to the larger amount of light emitted from the capillaries in Variant II array the corresponding fractions are 70% and 98%.

### 3.4 Determination of the point of interaction

In order to determine the pure resonant attenuation due to nitrogen in the inspected object in each capillary both on-resonance (protons) and off-resonance (electrons) events must be determined with the same position resolution. Thus, it is important to locate the point of interaction of the 9.17 MeV γ-ray in the detector for both types of events with similar precision. The proton track is short and its point of creation is determined with a precision of a single capillary. The Bragg peak, visible on the proton tracks can help locate the end of the track. Thus, the on-resonance point of interaction can be located with a precision of about 50 μm.

On the other hand, electrons have long and complex tracks without a visible Bragg distribution. Moreover, the start and end of a track is not so obvious. Several criteria for making a decision regarding the point of non-resonant interaction of the γ-ray have been devised:
- Complex tracks or events that produce several non-adjacent tracks are rejected.
- For electrons that escape the detector it is possible to locate the start position.
- The amount of energy deposited at the point of interaction is usually greater than at other positions along the track, due to the fact that the primary electron is strongly forward peaked and is thus moving along the length of the capillary. This phenomenon becomes more pronounced as the diameter of capillary increases.

By calculating the center-of-gravity for the simulated track deposited energy we obtain an estimate of point-of interaction.

We tested these criteria on 100 simulated electron events. Of these, 11% were rejected. Using the center-of-gravity for the remaining events, we have been successful in determining the interaction location of 85% to within a distance of 3 mm from the true point of interaction.

In conclusion, with the capillary array detector the GRA nitrogen image resolution has the potential of being 3×3 mm$^2$. As the number of electrons is much higher than that of the protons, the 11% rejection will not affect significantly the statistical precision of the nitrogen measurement.

### 4. Experimental

The experimental setup used in this work is based on the expertise and equipment derived from the CHORUS experiment and RD46 program at CERN [18-20]. The expertise accumulated at that time was recently recovered and reactivated at LHEP, University of Bern. In the following section we shall describe the various parts of the detector used for the proof-of principle experiment.

## 4.1 Capillary array unit

The capillary array unit is composed of ~20 µm diameter capillaries with wall thickness of ~11 µm, manufactured by Schott Fiber-Optics. The basic capillary array unit used here has a rectangular cross-sectional dimensions of 5×5 mm$^2$ as shown on Fig. 10 and was assembled into a larger area array. It was manufactured in lengths of ~1 meter and cut for our application to ~11 cm length.

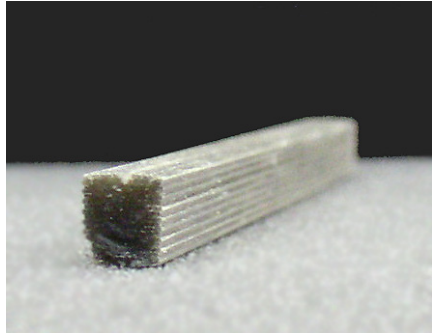

**Fig. 10: A capillary array unit used in the experiment**

The capillary matrix is made of Schott Technical Glass Type 8250, and incorporates extra mural absorbers (EMA) made to absorb light in order to prevent light cross-talk. The matrix glass has relatively low refractive index of 1.49. The capillaries are arranged in hexagonally shaped aggregates, also named "multies". These multies are then brought together in a beehive structure to create larger arrays of millimetric size. Fig. 11 shows a Scanning Electron Microscop (SEM) picture of the capillary array.

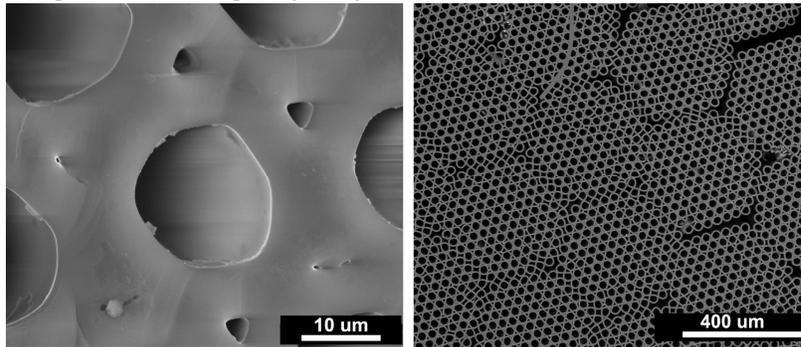

**Fig. 11: SEM images of capillary array used in the experiment. Scale: left-10 μm, right-400 μm.**

## 4.2 Liquid scintillator

The principal characteristics required of the new scintillator, in addition to high light yield, are: high refractive index, optical transparency, quench resistance and a strong solvent, capable of dissolving large concentrations of nitrogen-rich additives. These are very different properties from those of the scintillator currently used in the GRA application detectors.

The development and optimization of the special-purpose, nitrogen rich, high-refractive-index scintillator involves a tradeoff between the different requirements in order to achieve the composition that is most suitable for the application. This work will be described in detail in a separate publication.

For the purpose of experimental proof-of-principle a **non-nitrogenous** scintillator was prepared. It consists of 1-Methylnaphtalene (1MN), as a solvent, doped with 5g/l and 3/l of 2,5-Di-phenyl oxazole (PPO) and 1,4-bis-(2-methylstyryl)-benzene (bis-MSB) as primary and secondary phosphors respectively. Its properties are: refractive index-1.62, density-1.001 g/cc, light yield-77% of anthracene, peak emission wavelength-420 nm.

As mentioned in section 2.2 if a good quality mirror is used on the edge of the capillary array opposite to the optical readout, up to ~10% of the created light is expected to exit the capillary at the latter edge. In our experiment a reflecting mirror was not used.

### 4.3 Description of the detector

In order to construct a GRA detector, 4 square capillary array units depicted in Fig. 10 were glued together to form a larger 10 x 10 mm$^2$ assembly. This capillary assembly was placed inside an aluminum tube, and the space between them filled with silicon RTV elastosil E41 to prevent its filling with liquid scintillator. This material was chosen for its durability when exposed to the liquid scintillator.

Fig. 12 (left) illustrates a schematic diagram of the detector structure. The array is pressed against a Fiber Optic Plate (FOP) on the bottom side and a viewing glass window from the top, using aluminum flanges. The contact between the flanges and the tube is sealed using silicon gaskets. The filling inlet and outlet tubing are incorporated into the upper and lower flanges. During the filling procedure the upper and lower windows are shifted a short distance away from the array, to allow entry of the liquid into the capillaries. After the filling this gap is closed by tightening the plastic rod screws, such that both windows are in close contact with the capillary array. Fig. 12 (right) displays the assembled detector.

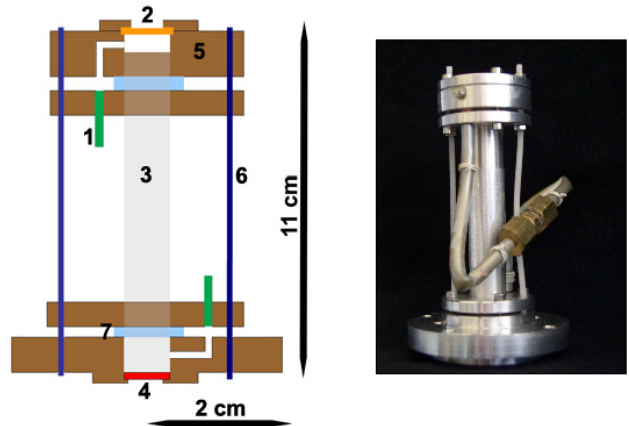

**Fig. 12: Detector schematic (left) and photograph (right). (1) liquid inlet/outlet with copper feed-through and plastic tubes; (2) glass window; (3) Aluminium tube housing for the capillary array, spaces filled with silicon RTV; (4) FOP; (5) aluminium flanges; (6) tightening plastic screw rods; (7) silicon gasket**

The connection to the filling system is via a copper feed-through and the plastic tubes seen in Fig. 12 - right which can connect to the filling system and to each other, creating a stand-alone detector cell.

### 4.4 Filling procedure

The filling of the capillaries is similar to the one used by the RD46 collaboration [20]. The filling system is described schematically on Fig. 13 (left). It consists of 2 glass reservoirs, a vacuum pump, pressurized Ar bottle and a series of valves. The system is first flushed with argon gas that eliminates the oxygen in the system and in the liquid scintillator stored in one of the vessels. The liquid scintillator is then introduced into the detector by applying argon pressure and the capillary array is repeatedly flushed with the liquid scintillator, which is shunt circularly between the two blue reservoirs. It takes several hours until all capillaries are filled.

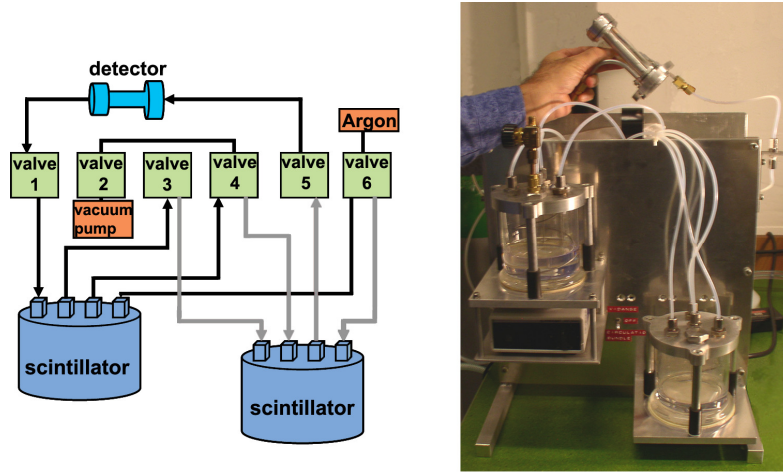

**Fig 13: Schematic diagram (left) and photo of the filling system**

The filling procedure is monitored by illuminating one end of the detector and observing the light emerging from the other end. For empty capillaries no light is internally reflected and they appear darker. Therefore, looking at a light source through the glass window of the capillary array detector reveals whether the filling process was successfully completed. Fig. 14 shows a photograph taken through the looking window of the filled detector.

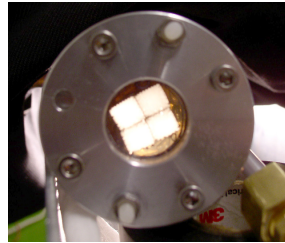

**Fig. 14: A successfully filled array**

Although this crude method cannot detect individual empty capillaries, it does give a good indication if the filling process was adequate for the requirements of this preliminary experiment.

### 4.5 Light readout system

The light readout was refurbished from the CHORUS system [20]. Fig. 15 is a schematic diagram of the readout system.

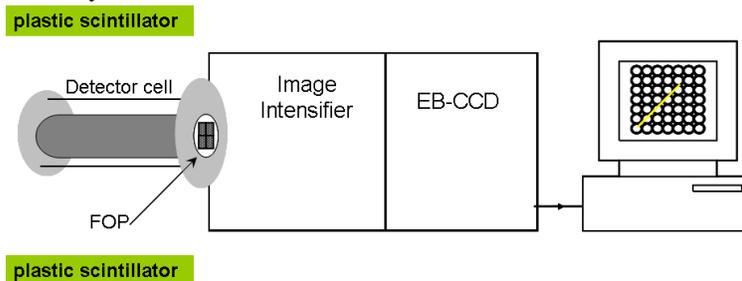

**Fig. 15: Schematic diagram of the readout system.**

The detector FOP was coupled to the entrance fiber-optics window of an electrostatically focused Image Intensifier (I-I) having a demagnification of $m_1=0.62$ and a green K-67 phosphor. The phosphor of the first I-I is viewed by an electron-bombarded-CCD (EBCCD) with an electrode gating, which combines the functions of an electrostatically-focused I-I and a conventional CCD. It is essentially an I-I with the phosphor screen replaced by a back-thinned CCD. The demagnification of the EBCCD is $m_2=0.83$ and it consists of 1024×1024 pixels, each with dimensions of 13×13 $\mu m^2$. The EBCCD output is displayed on a computer screen using a ROOT software interface. Both the I-I and the EBCCD were produced by Geosphera. The readout system was originally designed to be used with green-emitting (490-500 nm) scintillator so, for the blue-emitting scintillator employed here, its efficiency is lower by ~20%.

The detector was positioned between two plastic scintillator plates that served as cosmic-ray rejection system. Simultaneous signals in both plates provided an anticoincidence event veto which blocked the readout system.

### 4.6 Visualization of particle tracks

The experimental observation of proton and electron tracks was performed using a $^{239}$Pu-Be neutron source. The neutrons emitted from $^{239}$Pu-Be source have energy range of 0.2-10 MeV [25]. The average neutron energy is about 4.2 MeV [26]. The predominant γ-ray energy is 4.43 MeV, produced during the decay of the first excited state of $^{12}$C, populated in the $^9$Be($\alpha$,n)$^{12}$C$^*$ reaction.

During the experiment the readout system recorded the image on the capillary array face. The images were taken automatically with pre-set exposure gate times and intervals between exposures. The gates were in the range of 300 $\mu$s -100 ms. The lower limit on gate time was determined by the first I-I, the decay time of its phosphor being ~200 $\mu$s. The threshold for taking an event image was set so that the maximum numbers of tracks are visible while most of the noise (predominately single-pixel) is eliminated. Eventually, the optimal conditions were found, and the images were recorded with 1 ms gate and 500 mV threshold.

Fig. 16 shows the tracks obtained using the $^{239}$Pu-Be source. As the FOP was not in good physical contact with the back end of the array some of the liquid scintillator migrated into the gap. Thus, the light created in a single capillary expands as it traverses this scintillator, causing a "blooming" effect". This has the effect of delocalizing the pixel information and tends to blur the track images.

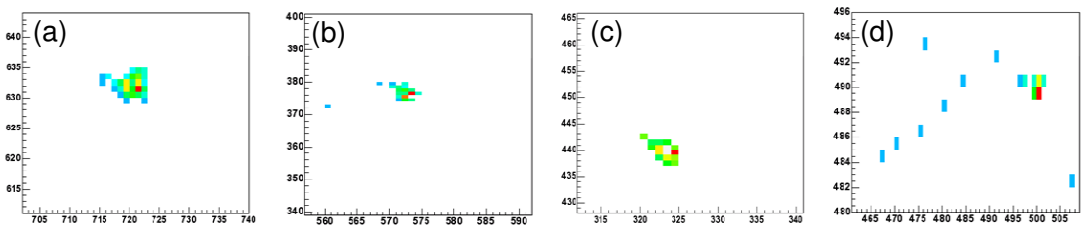

**Fig. 16: (a),(b),(c),(d) - tracks of recoil-protons produced by neutrons; (d) simultaneous events: high energy Compton electron and a recoil-proton.**

As seen in Fig. 16 (a)-(d), recoil-protons produced by neutrons give rise to bright, short tracks. In all the proton events displayed, there are indications of the Bragg peak, which manifests itself as a bright agglomeration of light near the end of the particle range.

## 5. Conclusions

This paper introduces a novel resonant detector for the GRA method with mm spatial resolution based on a micrometric glass capillary array filled with liquid scintillator. A detailed simulation of the response of the detector to electrons and protons generated by the 9.17 MeV γ-rays was followed by a proof-of principle experiment at LHEP, University of Bern, using a mixed γ-ray and neutron source. Towards this, a prototype capillary detector was assembled, including the associated filling and readout systems. Simulations and experimental results show that proton tracks are distinguishable from electron tracks at relevant energies, on the basis of a criterion that combines track length and light intensity per unit length.

The agreement between simulation and experiment is the crucial necessary (and, indeed, almost sufficient) condition for demonstrating the proof-of-principle in question. The sole remaining knowledge gap pertains to the optimization of the dedicated nitrogen-rich cocktail, an issue that all the commercial companies active in the field (as well as this group) have always solved satisfactorily and without undue effort. The development of a suitable high-refractive index, nitrogen-rich scintillator will be described in our next publication.

For the purpose of particle discrimination it is preferable to use larger capillaries which give rise to brighter electron tracks that could be analyzed with higher accuracy. It appears that a 50 μm diameter capillary with 5 μm walls represents a good compromise between resolution and light output.

As the proton track is short the on-resonance point of interaction can be located with a precision of about 50 μm, however the point of interaction of the non-resonant events can only be located with a resolution of 3×3 mm$^2$. This dictates the resolution of the nitrogen image. Thus, with the capillary array detector the nitrogen image resolution will be 3×3 mm$^2$. This compares very favorably with the 20×20 mm$^2$ resolution obtained with our present resonant detector.

In an operational GRA system operating with a 5 mA proton beam, the total event rate (electrons and protons) in a 20×20×240 mm$^3$ resonant detector positioned at about 250 cm from the $^{13}$C target, is expected to be about 500 events/s. Thus in order to obtain an image with not more than 2-3 particle tracks the exposure time of the readout system should be not more than about 6 ms/frame. It is then necessary for the image to be analysed or transferred, such that an automated track recognition procedure can be performed without introducing a significant dead-time or delay.

In summary, the advantages of the proposed detector are: Excellent separation between electrons and protons, improved spatial resolution (3×3 mm$^2$), possibility of constructing large area array detectors.

**Acknowledgments**

This work was funded by DHS Grant 071905-001. The authors would like to thank Drs Ronald Krauss and Curtis Bell from the Transportation Security Laboratory, Science and Technology Directorate, U.S. Department of Homeland Security**,** for their continuing interest in the GRA project.
Thanks are due to Drs. Guy Van Beek, Pierre Vilain and Gaston Wilquet of IIHE, the Inter-University Institute for High Energy Physics**,** ULB-VUB, Brussels, for many discussions and recommendations that proved critical for this research.